\documentclass[prd,twocolumn,superscriptaddress,preprintnumbers,balancelastpage,nofootinbib]{revtex4-1}

\usepackage{graphicx}  % needed for figures
\usepackage{dcolumn}   % needed for some tables
\usepackage{bm}        % for math
\usepackage{amssymb, amsmath}
\usepackage{slashed}   % for math

\usepackage[usenames,dvipsnames,svgnames,table]{xcolor}
\usepackage{comment}

\usepackage[utf8]{inputenc}
\renewcommand{\Re}{\mbox{Re}}

\def\lsim{\mathrel{\rlap{\lower4pt\hbox{\hskip1pt$\sim$}}
    \raise1pt\hbox{$<$}}}         %less than or approx. symbol
\def\gsim{\mathrel{\rlap{\lower4pt\hbox{\hskip1pt$\sim$}}
    \raise1pt\hbox{$>$}}}         %greater than or approx. symbol
\newcommand{\GeV}{{\rm GeV}}

\newcommand{\TeV}{{\rm TeV}}

\newcommand{\CP}{{\mathit  CP}}

\renewcommand{\thefootnote}{\fnsymbol{footnote}}
\def\beq#1\eeq{\begin{align}#1\end{align}}

\newcommand{\eg}{{\em e.g.}}

\RequirePackage{xspace}
\def\Bbar    {\kern 0.18em\overline{\kern -0.18em B}{}\xspace}

\def\Kbar    {\kern 0.18em\overline{\kern -0.18em K}{}\xspace}

\definecolor{BlueViolet}{rgb}{0.2, 0.00, 0.7}
\definecolor{Blue}{rgb}{0.15, 0.00, 0.9}
\usepackage[
colorlinks=true, linkcolor=Blue, citecolor=Blue, urlcolor=BlueViolet]{hyperref} 
\begin{document}
\widetext

%\preprint{}
\title{Implications for new physics from a novel puzzle 
in \boldmath{$\Bbar{}^0_{(s)} \to D^{(\ast)+}_{(s)} \{\pi^-,K^-\}$} decays}

\author{Syuhei Iguro}
\email{iguro@eken.phys.nagoya-u.ac.jp}
\affiliation{Department of Physics,
Nagoya University, Nagoya 464-8602, Japan}

\author{Teppei Kitahara}
\email{teppeik@kmi.nagoya-u.ac.jp}
\affiliation{Institute for Advanced Research, Nagoya University, Nagoya 464-8601, Japan}
\affiliation{Kobayashi-Maskawa Institute for the Origin of Particles and the Universe, 
Nagoya University,  Nagoya 464-8602, Japan}

%%%%%%%%%%%%%%%%%%%%%%%%%%%%%%%%%%%%%%%%
\begin{abstract}
\noindent
Recently, the standard model predictions for the $B$-meson hadronic decays, 
$\Bbar^0 \to D^{(\ast)+}K^-$ and 
$\Bbar^0_s \to D^{(\ast)+}_s \pi^-$, have been updated based on the QCD factorization approach. 
This improvement sheds light on a novel puzzle in the $B$-meson hadronic decays: there are mild but universal tensions between data and the predicted branching ratios.
Assuming the higher-order QCD corrections are not huge enough to solve the tension,
we examine several new physics interpretations of this puzzle.
We find that
the tension can be partially explained by a left-handed $W^\prime$ model, 
which can be compatible with other flavor observables and collider bounds.
\end{abstract}
%%%%%%%%%%%%%%%%%%%%%%%%%%%%%%%%%%%%%%%%

\maketitle

\renewcommand{\thefootnote}{\#\arabic{footnote}}
\setcounter{footnote}{0}

%%%%%%%%%%%%%%%%%%%%%%%%%%%%%%%%%%%%
\section{Introduction}
\label{sec:introduction}
%%%%%%%%%%%%%%%%%%%%%%%%%%%%%%%%%%%%

To test the standard model (SM) and search for physics beyond the SM,
precision measurements of meson decays, especially $B$-meson decays,
have been considerably investigated over the past 30 years.
In the meantime,
the experimental uncertainty has been surprisingly  reduced by experimentalists. 
On the other hand, theorists have 
played an equally important role: 
several approaches that can evaluate the QCD corrections have been invented, and 
the SM predictions have been sharpened.

Very recently, SM predictions for several $B$-meson hadronic decays are improved by Ref.~\cite{Bordone:2020gao}:
\begin{align}
\mathcal{B}(\Bbar^0 \to D^+ K^-)^{\rm exp}_{\rm SM} &=
  \left\{
  \begin{array}{l}
  \left( 1.86 \pm 0.20\right)\times 10^{-4}\,,\\
  \left( 3.26 \pm 0.15\right)\times 10^{-4}\,,
  \end{array}
  \right.
 \\
\mathcal{B}(\Bbar^0 \to D^{\ast +} K^-)^{\rm exp}_{\rm SM}&=
  \left\{
  \begin{array}{l}
  \left( 2.12 \pm 0.15\right) \times 10^{-4}\,,\\
  \left( 3.27 \,{}^{+0.39}_{-0.34}\right) \times 10^{-4}\,,
    \end{array}
  \right.
\\
\mathcal{B}(\Bbar_s^0 \to D^+_s \pi^-)^{\rm exp}_{\rm SM}&= 
  \left\{
  \begin{array}{l}
\left(3.00 \pm 0.23\right)\times 10^{-3}\,,\\
\left(4.42 \pm 0.21\right)\times 10^{-3}\,,
    \end{array}
  \right.
\\
\mathcal{B}(\Bbar_s^0 \to D^{\ast +}_s \pi^-)^{\rm exp}_{\rm SM}
& = 
  \left\{
  \begin{array}{l}
  \left(2.0 \pm 0.5\right)\times 10^{-3}\,,\\
  \left(4.3\, {}^{+0.9}_{-0.8}\right)\times 10^{-3}\,,
      \end{array}
  \right.
\end{align}
where the upper numbers are the PDG averages of the experimental data~\cite{PDG2020}, 
while the lower ones are the SM expectation values  
\cite{Bordone:2020gao}.
These SM predictions are obtained by the QCD factorization (QCDF) \cite{Beneke:1999br,Beneke:2000ry,Beneke:2001ev} at leading power in $\Lambda_{\rm QCD}/m_b$, where
the  Wilson coefficients at next-to-next-to-leading logarithmic accuracy are used \cite{Gorbahn:2004my}. 
Compared to the previous estimations  \cite{Huber:2016xod}, the theoretical uncertainties are significantly reduced 
thanks to recent developments in the $\Bbar_{(s)} \to D_{(s)}^{(\ast)}$ form factors including order
$\mathcal{O}(1/m_c^2)$ corrections within the framework of the heavy-quark expansion
\cite{Jung:2018lfu,Bordone:2019vic,Bordone:2019guc,Iguro:2020cpg}.

These hadronic channels are theoretically clean 
due to the absence of
 penguin and annihilation topologies. 
Furthermore, 
resultant amplitudes are dominated by the color-favored tree topology.

Above SM predictions deviate from the data at 
$5.6\sigma$ ($D^+K^-$), $3.1\sigma$ ($D^{\ast +}K^-$), $4.6\sigma$ ($D^+_s \pi^-$), and $2.4\sigma$  levels  ($D^{\ast +}_s \pi^-$), respectively.
Surprisingly, all deviations are in the same direction and  similar size.
Note that 
$\mathcal{B}(\Bbar^0 \to D^+ \pi^-)_{\rm SM} = (3.93 {}^{+0.43}_{
-0.42})\times 10^{-3}$ and 
$\mathcal{B}(\Bbar^0 \to D^{\ast +} \pi^-)_{\rm SM} = (3.45 {}^{+0.53}_{
-0.50})\times 10^{-3}$, which are evaluated in Ref.~\cite{Huber:2016xod}, 
also deviate from the data, 
$\mathcal{B}(\Bbar^0 \to D^+ \pi^-)^{\rm exp} = (2.52 \pm  0.13)\times 10^{-3}$ and 
$\mathcal{B}(\Bbar^0 \to D^{\ast +} \pi^-)^{\rm exp} = (2.74 \pm 0.13)\times 10^{-3}$~\cite{PDG2020} at the $3.2\sigma$ and $1.4\sigma$ levels, respectively.

Within the SM,
there are two possibilities that these tensions are alleviated.
The first possibility is an input value of $|V_{cb}|$.
For $|V_{cb}|$, the authors of Ref.~\cite{Bordone:2020gao} use an average of the inclusive and exclusive  determinations in the $B$-meson semileptonic decays:
$|V_{cb}|=(41.1 \pm 0.5)\times 10^{-3}$
\cite{Bordone:2019vic,Bordone:2019guc}.
If one adopts the exclusive $|V_{cb}|$, $|V_{cb}|=(39.25 \pm 0.56)\times 10^{-3}$ \cite{Amhis:2019ckw},
amplitudes of the above processes are uniformly reduced by $4.5\%$.
Note that the exclusive $|V_{cb}|$, however, produces an additional $4.2\sigma$ level tension in $\varepsilon_K$ \cite{Kim:2019vic}.
See also for a more recent determination of the exclusive $|V_{cb}|$ using the full angular distribution data \cite{Iguro:2020cpg}. 

Another possibility is higher-order QCD corrections.
The next-to-leading power and next-to-next-to-leading power corrections to the QCDF amplitudes
are also estimated by the same authors \cite{Bordone:2020gao}, and the sizes of 
those corrections to the amplitudes are evaluated as 
$\mathcal{O}(1)\%$.

The above puzzled situation could be resolved by introducing new physics contributions to $b \to c \bar{u}q$ transitions, where $q = d$ and $s$.
Furthermore, it is shown
that all ratios between these branching fractions are consistent with data \cite{Bordone:2020gao}.
It clearly implies that the new physics effects should be universal in $b \to c \bar{u}q$ transitions.
Therefore, the following questions are interesting: 
whether such a new physics is still allowed by the other flavor constraints and by the hadron collider constraints, and
how much the tensions can be alleviated by a valid new physics model.
Below we will refer to this puzzle as $b\to c \bar{u}q$ anomaly.
In this Letter,
we examine several new physics scenarios to explain the $b\to c \bar{u}q$ anomaly.

%%%%%%%%%%%%%%%%%%%%%%%%%%%%%%%%%%%%
\section{Framework}
\label{sec:frame}
%%%%%%%%%%%%%%%%%%%%%%%%%%%%%%%%%%%%

We consider the following effective Lagrangian to investigate new physics contributions to $b\to c\bar{u}q$ processes:
\begin{align}
\mathcal{L} = - \frac{4 G_F}{ \sqrt{2}} \sum_q V_{cb} V_{uq}^\ast
\sum_{i=1,2} C_i^q (\mu) \mathcal{Q}_i^q  (\mu) \,,
\label{eq:Leff}
\end{align}
with  the left-handed current-current operators in the CMM basis \cite{Chetyrkin:1996vx,Chetyrkin:1997gb},
\begin{align}
\mathcal{Q}_1^q &= (\bar{c}_L \gamma^{\mu} T^a b_L) (\bar{q}_L \gamma_{\mu} T^a  u_L)\,,\\
\mathcal{Q}_2^q &= (\bar{c}_L \gamma^{\mu}  b_L) (\bar{q}_L \gamma_{\mu}  u_L)\,,
\end{align}
where $q=d,s$. $T^a$ is the SU(3$)_C$ generator, and
$V$ is the Cabibbo-Kobayashi-Maskawa matrix \cite{Cabibbo:1963yz,Kobayashi:1973fv}. 
In our analysis, we refrain from adding operators that are absent in the SM, \eg, 
$(\bar{c}_L b_R)(\bar{q}_L u_R)$. 
We will discuss this possibility in the last section.

New physics contributions to the Wilson coefficients, 
$C_1^{q,{\rm NP}}$ and $C_2^{q,{\rm NP}}$, 
become involved at the new physics scale $\Lambda$.
These values are modified by the renormalization-group (RG) evolution 
from $\Lambda$ down to the hadronic scale $m_b$.
The leading-order (LO) QCD RG evolution is summarized in 
Appendix~\ref{App:RG}.
For instance, when $\Lambda = 1\,\TeV$, we obtain  an evolution matrix as
\begin{align}
\begin{pmatrix}
C_1^{\rm NP} (m_b) \\
C_2^{\rm NP}  (m_b)
\end{pmatrix}
= 
\begin{pmatrix}
1.36 & 
-0.87 
\\
-0.19   & 1.07
\end{pmatrix} \begin{pmatrix}
C_1^{\rm NP}  (1\,\TeV) \\
C_2^{\rm NP}  (1\,\TeV)
\end{pmatrix}
\,.
\end{align}

It is found that 
a universal destructive shift in the SM contributions is favored in the $b\to c\bar{u}q$ anomaly \cite{Bordone:2020gao}.
The preferred size is $\sim - 17\%$, which corresponds to $C_2^{d,{\rm NP}}=C_2^{s,{\rm NP}}=C_2^{\rm NP}$ and 
\begin{align}
\frac{C_2^{\rm NP}(m_b)}{C_2^{\rm SM}(m_b) 
}  
= -0.17 \pm 0.03\,.
\label{eq:NPrequired}
\end{align}

It is checked that such a new physics contribution is compatible with data of
the total decay rates of the $B$-mesons, $\tau_{B_s}/\tau_{B_d}$, and $a_d^{fs}$ \cite{Bordone:2020gao,Bobeth:2014rda,Brod:2014bfa,Lenz:2019lvd}.
Another potentially strong constraint comes from the kaon hadronic decays ($s \to u \bar{u} d$).
The $\CP$-conserving parts of the isospin amplitudes, $A_{I}=\langle (\pi \pi )_{I} |\mathcal{H}^{|\Delta S| =1}_{\rm eff}|K \rangle $ for $I=0,2$, have been measured very precisely through all $K \to \pi \pi$ data \cite{Kitahara:2016nld,Blum:2015ywa}
\begin{align}
\Re A_0^{\rm exp} &= \left(3.3201 \pm 0.0018\right) \times 10^{-7}\,\GeV\,,\\
\Re A_2^{\rm exp} &= \left(1.4787 \pm 0.0031\right) \times 10^{-8}\,\GeV\,.
\end{align}
On the other hand,
these theoretical predictions
are
\begin{align}
\Re A_0^{\rm SM} &= \left(2.99 \pm 0.67\right) \times 10^{-7}\,\GeV\,,\\
\Re A_2^{\rm SM} &= \left(1.50 \pm 0.15 \right) \times 10^{-8}\,\GeV\,,
\label{eq:Kpipi}
\end{align}
where the hadronic matrix elements are calculated by the lattice QCD simulations \cite{Blum:2011ng,Blum:2015ywa, Bai:2015nea,Ishizuka:2018qbn,Abbott:2020hxn}.
Although the $A_2$ amplitude is more sensitive to new physics than $A_0$, 
we find that a $\pm 20\%$ new physics contribution to the $s \to u \bar{u} d$ amplitude could be compatible with the data.

%%%%%%%%%%%%%%%%%%%%%%%%%%%%%%%%%%%%
\section{Minimal flavor violation}
\label{sec:MFV}
%%%%%%%%%%%%%%%%%%%%%%%%%%%%%%%%%%%%

First, we study the most simple possibility for new physics scenario:
minimal flavor violation (MFV) hypothesis 
\cite{DAmbrosio:2002vsn,Isidori:2012ts}.
The detailed calculations for this section can be found in Appendix~\ref{sec:MFV}.

We examine a dimension-six operator,
$\mathcal{L} =  1/(2 \Lambda^2)  ( \bar{Q}_L  \gamma^{\mu}  Q_L )^2$, whose flavor off-diagonal components are controlled by the quark Yukawa.
In the quark mass-diagonal basis,
this operator produces
$C_{2}^{q,{\rm MFV}} (\Lambda)\sim  -  1/ (2 \sqrt{2} G_F \Lambda^2) $.
Then, the $b\to c \bar{u} q$ anomaly in Eq.~\eqref{eq:NPrequired} suggests  $\Lambda \lesssim 0.49\,\TeV$.

Among the various flavor and collider constraints, a nonresonant dijet angular distribution search in the LHC gives the most stringent constraint on this scenario.
The result is reported by the ATLAS collaboration 
at $\sqrt{s} = 13\,\TeV$ with $\int dt \mathcal{L}=37$\,fb$^{-1}$ \cite{Aaboud:2017yvp}.
We interpret the result and obtain a $95\%$ C.L. exclusion limit as
$\Lambda < 3.7\,\TeV $, which excludes the suggested $\Lambda \sim 0.49\,\TeV$.
From this collider constraint, we obtain a bound
\begin{align}
\frac{C_2^{\rm MFV}(m_b)}{C_2^{\rm SM}(m_b)}  \gtrsim -0.002\,.
\end{align}
Hence, this %new physics
scenario never explains the~$b\to c \bar{u}q$~anomaly.

%%%%%%%%%%%%%%%%%%%%%%%%%%%%%%%%%%%%
\section{SU(2)$\times$SU(2)$\times$U(1) model}
\label{sec:SU2}
%%%%%%%%%%%%%%%%%%%%%%%%%%%%%%%%%%%%

Next,
we consider a new physics model that can produce a more convoluted flavor structure.
An extended electroweak gauge group 
SU(2$)_1\times$SU(2$)_2\times$U(1$)_Y$
with heavy vectorlike fermions
produces heavy gauge bosons, $W^{\prime \pm}$ and $Z^\prime$,
interacting with the left-handed SM fermions with a nontrivial flavor structure \cite{Langacker:1988ur,Chivukula:2003wj,Chiang:2009kb,Boucenna:2016wpr,Boucenna:2016qad}.
These flavor structures are controlled by
the number of generations of the vectorlike fermions ($n_{\rm VF}$)
and mixings between the SM fermions and vectorlike fermions.

The  heavy gauge boson interactions are  \cite{Boucenna:2016qad}
\begin{align}
\mathcal{L}=&+ \frac{g_{ij}}{2}Z^\prime_\mu \bar{d}_L^i \gamma^\mu d_L^j 
- \frac{\left(V g V^\dag \right)_{ij}}{2} Z^\prime_\mu \bar{u}_L^i
\gamma^\mu  u_L^j\nonumber\\
& -\frac{\left(V g\right)_{ij}}{\sqrt{2}} W^{\prime +}_\mu \bar{u}_L^i \gamma^\mu  d_L^j+
{\rm H.c.}\,,
\end{align}
where $u_L, d_L$ are the mass eigenstates, and 
a coupling $g_{ij}$ is defined in the $d_L$ basis.
In the following, we will take $M_{W'} = M_{Z'}=M_V$ for simplicity.
By integrating out $W^{\prime \pm}$, new physics contribution $C_2^{q,{W'}}$ is obtained  as 
\begin{align}
C_2^{q,W'}(M_V)
=& \frac{1}{4\sqrt{2} G_F M_V^2}\frac{(Vg)_{23} (Vg)_{1q}^\ast }{V_{cb}V_{uq}^\ast }\,.
\label{C2Wp}
\end{align}

In order to generate an uniform shift
in both $b\to c \bar{u} d$ and $b\to c \bar{u} s$, 
a SM-like flavor structure in $(Vg)_{1q}$ is required,
and hence $g_{11}$ should be nonzero.
When only $g_{11}$ is a nonzero entry in $g_{ij}$,
a dangerous $\bar{c}u Z'$ flavor-changing neutral current
 is generated and 
it is severely~constrained by the $D$-meson mixing as 
$|g_{11}|/M_V < {\mathcal{O}}(10^{-2})$\,(TeV)$^{-1}$ \cite{Golowich:2007ka}.
To evade this bound, we follow the U(2$)^3$ flavor symmetry \cite{Barbieri:1995uv,Barbieri:2011ci} 
and take $g_{11}=g_{22}$  in $g_{ij}$ in the following analyses.
Then the bound from the $D$-meson mixing is significantly relaxed as 
$
|g_{11}|/M_V \lesssim 16 \left( \rm{TeV} \right)^{-1}.
$

Another flavor constraint comes from the $K\to\pi\pi$ data. 
By permitting a $\pm 20\%$ new physics contribution to the Wilson coefficient of $(\bar{u}_L \gamma^{\mu} d_L)(\bar{s}_L \gamma_{\mu} u_L)$ [see Eq.~\eqref{eq:Kpipi}],
we obtain 
\begin{align}
|g_{11}|/M_V \lesssim 3.6 \left( \rm{TeV} \right)^{-1}\,.
\label{Kpipi}
\end{align}
Note that  many types of diagrams contribute to $K\to\pi\pi$ decays,
and nonperturbative QCD plays an essential role there. 
Therefore, this bound is a just reference value.

In addition to $g_{11}$,
another nonzero entry of $g_{33}$ or $g_{23}$ is necessary to produce $C_2^{q, W'}$.
Therefore, we consider the following flavor texture:
\begin{align}
g_{ij} = 
\begin{pmatrix}
g_{11} & 0 & 0 \\
0 & g_{11} & g_{23}\\
0 & g_{23}  & g_{33}
\end{pmatrix}
\,,
\label{eq:gij}
\end{align}
and will discuss several scenarios in detail. 
We assume $g_{ij}$ is real for simplicity.
Note that 
when $g_{11}$ is $\mathcal{O}(1)$,
production cross sections of the heavy gauge bosons 
become considerably large in the hadron collider,
and hence we will mostly discuss the LHC constraints in each subsection.
To evade surveying a dedicated collider constraint for low-mass region where the constraint would be more stringent, 
the mass range $M_V> 1\,\TeV$ is considered in our analysis.

\vspace{-0.2cm}
\subsection{Scenario 1: \boldmath{$g_{11}$ and $g_{33}$}}
\label{SU2:S1}

In this subsection, 
we take $g_{23}=0$ and consider a scenario of
$g_{ij} = \textrm{diag}(g_{11},\,g_{11},\,g_{33})$.
Such a flavor structure can be obtained from $n_{\rm VF}=1$.
In this case,
$(Vg)_{23}$ in Eq.~\eqref{C2Wp} comes from $V_{cb}g_{33}$.
Since one has a factor of $V_{cb}$
just as the SM,
$\sqrt{ | g_{11}g_{33}| }/M_V$ should be larger than  $\mathcal{O}(1)\,\TeV^{-1}$ to generate 
 new physics contributions to $b \to c \bar{u} q$ processes (see  previous section).
Furthermore, a relative sign between $g_{11}$ and $g_{33}$ must be negative to produce the destructive interference with the SM in the $b\to c \bar{u}q$ decays.
A requirement of the $b\to c \bar{u} q$ anomaly
within 2\,$\sigma$ level leads to 
\begin{align}
2.6  \,(\TeV)^{-1} \lesssim \sqrt{|g_{11}g_{33}|}  /M_V \lesssim 3.8 \,(\TeV)^{-1}\,.
\label{eq:anomaly_scenario1}
\end{align}
Therefore, large couplings are necessary in this scenario.

First, let us examine the constraint from the $B_{s}$-meson mass difference ($\Delta M_{{s}}$).
In this scenario, the dominant contribution comes from a $W$--$W'$ box diagram.
We observed that the GIM mechanism still works in this flavor structure, 
and obtain a simple formula for the $W$--$W'$ box contribution to $\Delta M_s$, 
\begin{align}
\frac{\Delta M_{{s}}^{W^\prime}}{\Delta M_{{s}}^{\rm SM}} \simeq 
\eta^{\frac{2}{7}} \frac{2 g_{11}g_{33}f^\prime(x_t,x_{V})}{g_W^2 f(x_t)} \,, 
\label{WWpBox}
\end{align}
with $\eta = \alpha_s(M_V)/\alpha_s(m_W)$, 
$x_t=m_t^2/m_W^2$ and $x_{V}=M_V^2/m_W^2$, and 
$g_W$ is the weak coupling.
The loop functions are defined in Appendix~\ref{App:loop}.
We also have the same shift in $B_d$-meson mixing,
but it is less constrained because of its large theoretical uncertainty.  
By imposing that 
the new physics contribution is within $2\sigma$ uncertainty of  $\Delta M_{{s}}^{\rm SM}$ \cite{Blanke:2018cya,DiLuzio:2019jyq},   
we~obtain
\begin{align}
\sqrt{| g_{11}g_{33}| }/M_V \lesssim 1.7 \left( \rm{TeV} \right)^{-1}\,.
\label{eq:deltaMs}
\end{align}
Although 
$\Delta M_s$ bound is incompatible with the $b\to c \bar{u}q$ anomaly in Eq.~\eqref{eq:anomaly_scenario1},
we want to know how much this scenario can 
alleviate
the puzzle.

%%%%%%%%%%%%%%%%%%%
\begin{figure*}[t]
\begin{center}
 \includegraphics[width=0.32\textwidth]{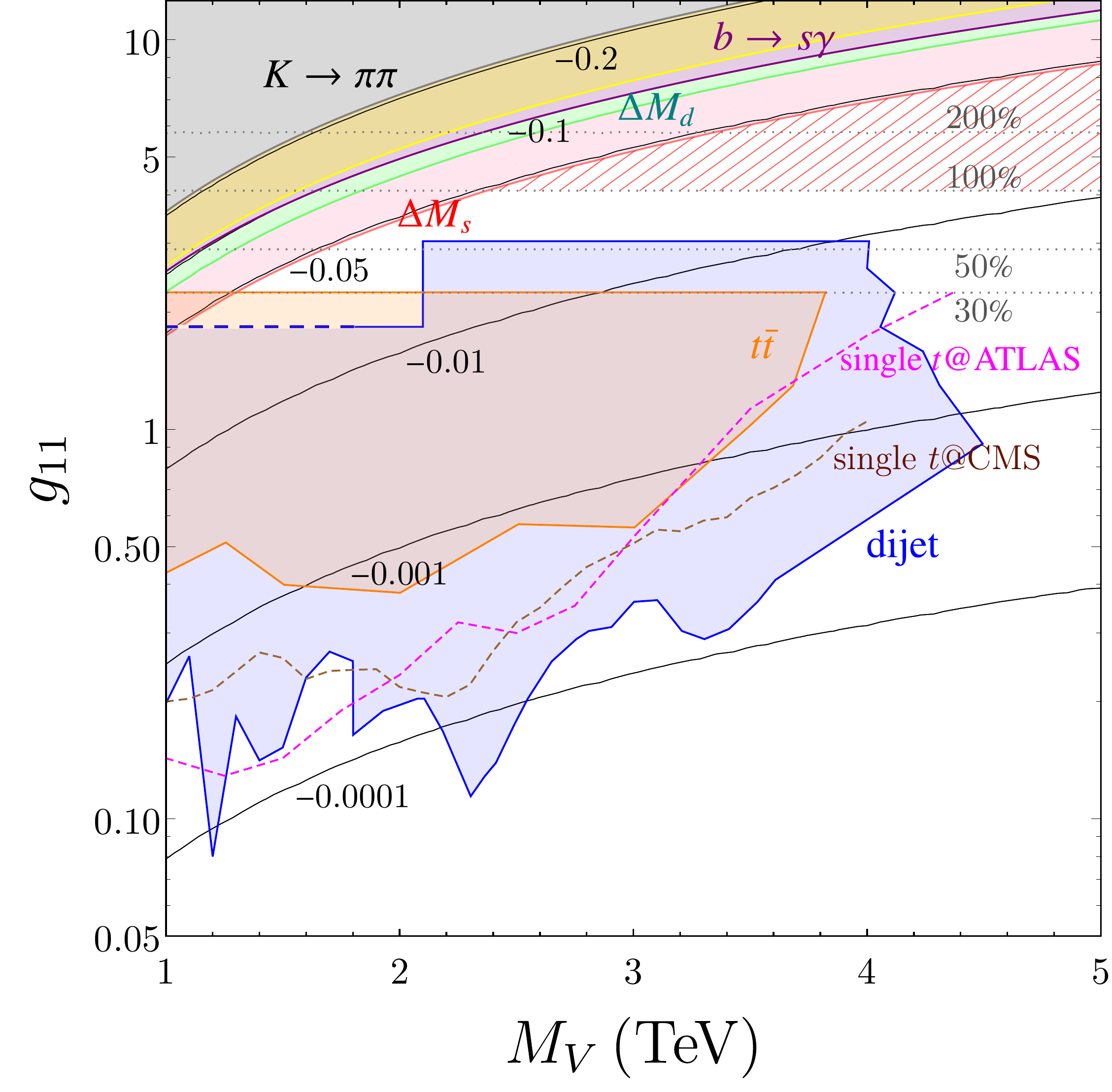}
~~%
 \includegraphics[width=0.32\textwidth]{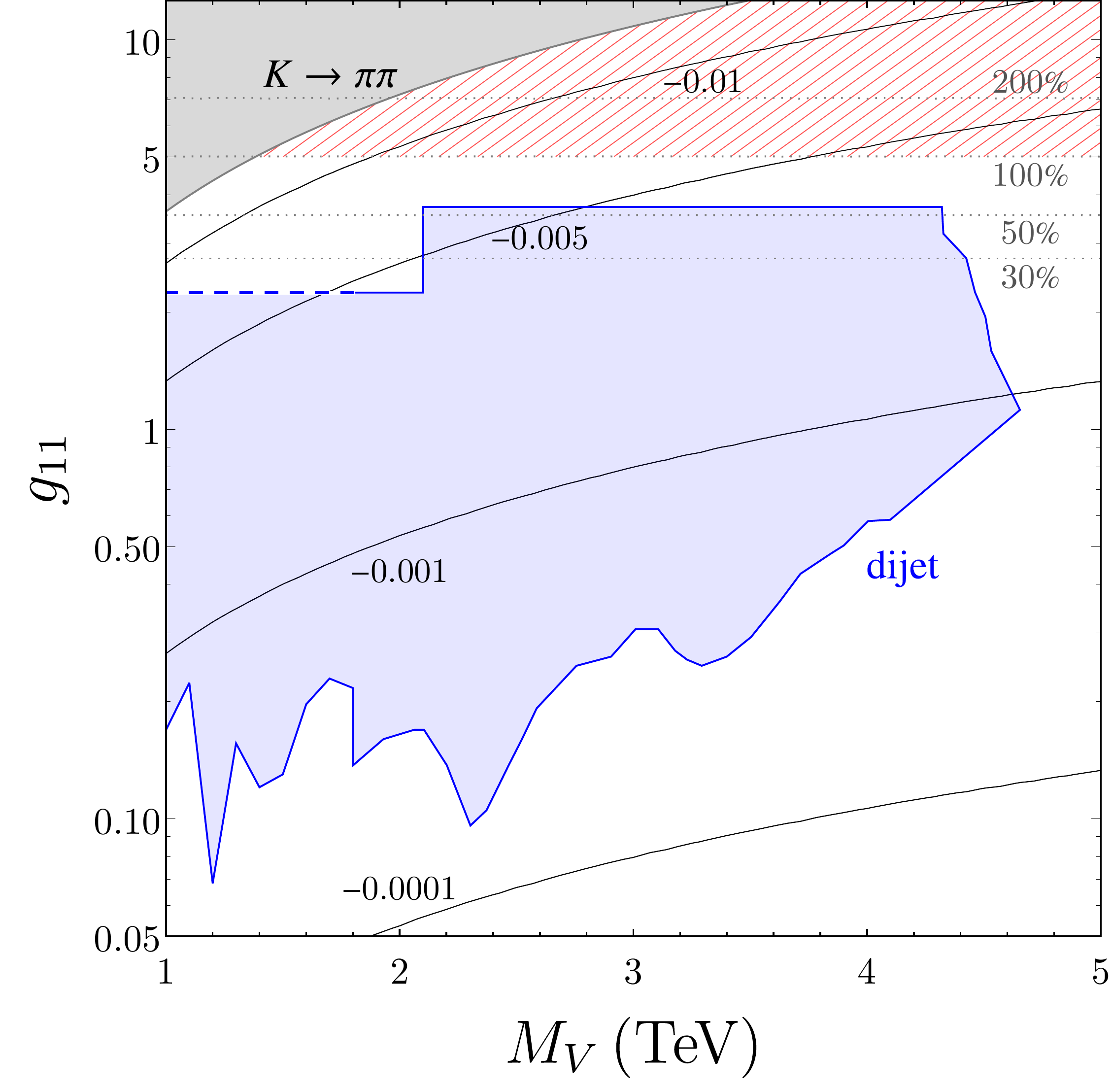}
~~%
 \includegraphics[width=0.32\textwidth]{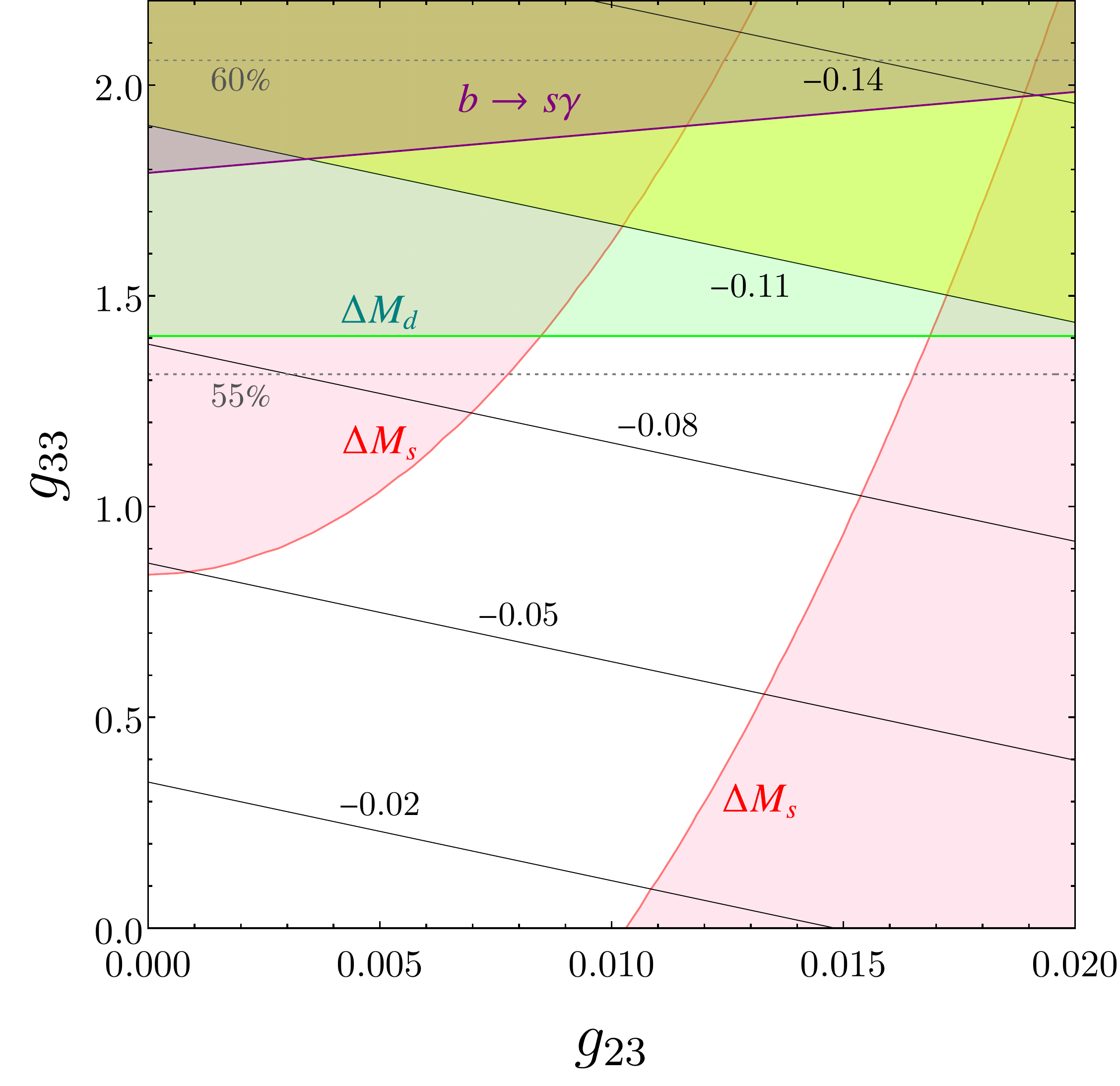}
     \vspace{.1cm}
\caption{
Contours of $C_2^{\rm NP}(m_b)/C_2^{\rm SM}(m_b) $ are presented. The puzzle can be explained at $2\sigma$ level in the yellow bands.
        The blue and orange shaded regions are excluded by the dijet  \cite{Aad:2019hjw,Sirunyan:2019vgj,Sirunyan:2018xlo} and $t\bar{t}$ searches \cite{Sirunyan:2018ryr,Aad:2020kop}  at $95\%$ C.L., respectively.
        The regions above the dashed lines are excluded by the single $t$ searches in the NWA (see text) \cite{Sirunyan:2017vkm,Aaboud:2018jux}.
       Furthermore,   the gray, red, green, and 
         purple shaded regions are constrained by $K\to \pi \pi$, $\Delta M_s$, $\Delta M_d$, and $b \to  s \gamma$,  respectively.
         The dotted line indicates $\Gamma_V /m_V$ and the red-hatched regions represent $\Gamma_V /m_V > 100\%$.
        {\bf Left:} scenario 1. We take $g_{33} = - g_{11}$.
        {\bf Middle:}  scenario 2. We take $g_{23} = - 0.01 (M_V/\TeV)$.
        {\bf Right:} scenario 3.
        We take  $M_V=1\,\TeV$
 and $g_{11}= - 3.6$. 
        }
    \label{fig}
    \vspace{.0cm}
\end{center}
\end{figure*}
%%%%%%%%%%%%%%%%%%%%%

Next, we consider constraints from resonant productions of the heavy gauge bosons at the LHC.
When $g_{11}$ and $g_{33}$ entries are nonzero,
$Z^\prime$ is produced via $pp\to q \bar{q}\to Z^\prime$ and  also $pp\to b\bar{b} \to Z^\prime$,
while 
$W^{\prime \pm}$ is produced thorough $pp\to q \bar{q}' \to W^{\prime \pm}$ processes.
When $M_V\gg m_t$, the decay width of those particle is approximately given as,
\begin{align}
\Gamma_{V=W^{\prime},Z^\prime}
\simeq \frac{2|g_{11}|^2+|g_{33}|^2}{16\pi}m_{V}\,.
\end{align}
We find that relevant collider bounds come from 
dijet and $t\bar{t}$ searches.
The former provides the relevant bound for $|g_{11}| \gg |g_{33}|$, while the latter for $|g_{11}| \lsim |g_{33}|$.

Currently, both ATLAS and CMS collaborations 
 reported upper limits 
on the heavy dijet resonance cross section
using the data of $\sim140 \,{\rm fb}^{-1}$\,\cite{Aad:2019hjw,Sirunyan:2019vgj}.
Since $\mathcal{O}$(1) couplings are necessary to relax the tension, the decay width can be not small.
Therefore,
we adopt width-dependent limits 
on the cross section times the dijet branching ratio.
The broader the width is, the weaker the limits become 
because a characteristic resonance peak is diluted.
The search is robust up to $\Gamma_V/M_V =20\%$ for $1.8$--$2.1\,\TeV$, and 
up to $\Gamma_V / m_V =55\%$ for the heavier region \cite{Sirunyan:2019vgj}.
For the mass range of $1$--$1.8\,\TeV$, 
we use an upper limit in Ref.~\cite{Sirunyan:2018xlo},
where the narrow width approximation (NWA) is used.
As for the heavy $t\bar{t}$ resonance search,
CMS reported the width-dependent limit using the data of 36\,${\rm fb}^{-1}$ up to $\Gamma_V/M_V =30\%$ \cite{Sirunyan:2018ryr},
while
ATLAS reported the result using the data of 139\,${\rm fb}^{-1}$ in the NWA \cite{Aad:2020kop}.

We obtained the production cross section of $Z^\prime$ and $W^{\prime \pm}$ by rescaling the result in Refs.~\cite{Sirunyan:2019vgj, Hayreter:2019dzc}, where 
$\sigma(pp \to q\bar{q}' \to W^{\prime +}) + \sigma(pp\to q\bar{q}'  \to W^{\prime -}) \simeq  2  \sigma(pp\to q\bar{q} \to Z^{\prime})$ is used \cite{Abe:2015uaa}.
The excluded regions from the dijet and $t\bar{t}$ searches 
are shown as the blue and orange shaded regions in Fig.~\ref{fig} (left), 
respectively.

We also show constraints from the single $t$ searches by using the data of $\sim 36\,$fb$^{-1}$ of CMS \cite{Sirunyan:2017vkm} and ATLAS~\cite{Aaboud:2018jux}:
the regions above the dashed lines  in Fig.~\ref{fig} (left) are excluded. 
Note that both analyses assume the narrow resonance, and no study exists for broad resonances.

Taking a conservative position, 
regions above the plateaus of the shaded areas 
can not be excluded,
where the corresponding $\Gamma_V/M_V$ exceeds the maximum width 
shown in each experimental result: $\Gamma_V/M_V>30\%$ in the $t \bar{t}$ search,
 and $\Gamma_V/M_V > 55 \%$ for $2.1$--$5\,\TeV$ and  $\Gamma_V/M_V>20\%$  for  $1.8$--$2.1\,\TeV$ in the dijet search.
The horizontal blue dashed lines are extrapolations obtained 
by assuming the analysis of Ref.~\cite{Sirunyan:2018xlo} is valid up to $\Gamma_V/M_V=20\%$, 
and should be taken with more care. 
We note that limits from the dijet angular 
distribution data, which are not considered here, 
would also depend on the width-mass ratio 
and only contact interaction models are investigated \cite{ATLAS:2012pu,Aaboud:2017yvp}.
Further dedicated analysis would be necessary to exclude such a broad width region.

The red-hatched regions represent $\Gamma_V > m_V$, where a particle picture is no longer valid and one could not discuss any conclusive prediction.
%\\
%$~\,\,\,\,\,$ 

Note that both our study and above experimental analyses have considered only the $s$-channel productions of $W^\prime$ and $Z^\prime$, although there are several $t$-channel contributions.
Since the $t$-channel process does not show a resonant nature, and there is a huge QCD $t$-channel background in the dijet production,
we  suppose that inclusion of the $t$-channel processes in the signal  could not amplify the signal-to-noise ratio in the resonance searches.
Such $t$-channel contributions, which are insensitive to the width, would be potentially accessible in the angular distribution search.

%\\
%$~\,\,\,\,$ 

As long as we allow the broad width scenario,
we find that the bound from $\Delta M_{s}$ in Eq.~\eqref{eq:deltaMs} determines the maximal deviation of $C_2^{W^\prime}/C_2^{\rm SM}$, which is 
independent of the ratio of $g_{11}$ and $g_{33}$.
For these reasons, we conclude $C_2^{W^\prime}/C_2^{\rm SM} \gtrsim -0.05$ when 
 $g_{23}=0$.

\subsection{Scenario 2: \boldmath{$g_{11}$ and $g_{23}$}}
\label{SU2:S2}

For the second scenario,
we set $g_{33}=0$ and consider $g_{11}$ and $g_{23}$  in Eq.~\eqref{eq:gij}.
Such a flavor structure can be obtained when $n_{\rm VF}=2$.
In this scenario, the $b\to c\bar{u}q $ anomaly requires 
\begin{align}
0.54 \,(\TeV)^{-1}  \lesssim \sqrt{|g_{11}g_{23}|}/M_V  \lesssim 0.78  \,(\TeV)^{-1} \,.
\label{eq:anomalyscenario2}
\end{align}
Although the size of the required coupling product is much smaller than the previous scenario, 
a severe bound on $g_{23}$ comes from $\Delta M_{s}$,
where there is a tree-level $Z^\prime$ exchange diagram.
We obtain 
\begin{align}
\frac{\Delta M_{{s}}^{Z^\prime}}{\Delta M_{{s}}^{\rm SM}} \simeq
\eta^{\frac{2}{7}} \frac{16\pi^2 g_{23}^2}{ g_W^4 (V_{ts}V_{tb}^\ast)^2 x_V x_t f(x_t)} \,, 
\label{ZpBs}
\end{align}
and find that $g_{23}$ always gives a positive shift in $\Delta M_{{s}}$.
The constraint from $\Delta M_{{s}}$ is
 \cite{Blanke:2018cya,DiLuzio:2019jyq}
\begin{align}
|g_{23}|/M_V  \lesssim 0.01 \,(\TeV)^{-1}\,.
\label{eq:Ms2}
\end{align}

Therefore, $g_{11} \gtrsim 30 \, (M_V/\TeV) \gg 4 \pi$ is required by  Eqs.~\eqref{eq:anomalyscenario2} and \eqref{eq:Ms2}, which implies that the $b\to c\bar{u}q$ anomaly can not be explained by this scenario.

In this scenario, $|g_{23}|\ll |g_{11}|$ should be satisfied.
This simplifies the collider constraints
because the production cross section 
is controlled only by $|g_{11}|$,
and the heavy gauge bosons decay into jets with $\mathcal{B} \simeq 1$.
The constraints are shown in Fig.~\ref{fig} (middle).
We find
$C_2^{W^\prime}/C_2^{\rm SM}  \gtrsim - 0.01$, where $g_{23} = - 0.01 (M_V/\TeV)$ is taken.

\subsection{Scenario 3: \boldmath{$g_{11}$, $g_{23}$ and $g_{33}$}}
\label{SU2:S3}

To see maximum value of $|C_2^{W^\prime}/C_2^{\rm SM}|$ in this model,
we combine the first and second scenarios:
all $g_{11}$, $g_{23}$, and $g_{33}$ are non-zero entries.
The point of this scenario is that the severe bound from $\Delta M_s$ can be turned off by
\begin{align}
\frac{\Delta M_{{s}}^{W^\prime}}{\Delta M_{{s}}^{\rm SM}}
+ \frac{\Delta M_{{s}}^{Z^\prime}}{\Delta M_{{s}}^{\rm SM}} \sim 0\,,
\end{align}
where the $W^\prime$ contribution is destructive and the $Z^\prime$ one is constructive in $\Delta M_s$ (see previous subsections).
We find, however, that 
even if the $\Delta M_{s}$ bound is turned off, 
 $g_{11}g_{33}$ is still constrained  from the  $\Delta M_{d}$ as
\begin{align}
\sqrt{| g_{11}g_{33}| }/M_V \lesssim 2.3\, \left( \rm{TeV} \right)^{-1}\,.
\end{align}
This bound restricts the possible $W^\prime$ contribution to the $b\to c\bar{u}q$ processes.
Also, we have checked a constraint from $b \to s \gamma $ data. We conclude that the  $b \to s \gamma $ bound is less sensitive than $\Delta M_{d}$, see Appendix~\ref{App:bsg}.

Since $|g_{23}| \ll |g_{11}|, |g_{33}|$ still holds in this scenario,
the collider constraints are almost the same as the scenario~1.
We focus on a parameter region that the all LHC constrains are evaded by the broad width of the heavy gauge bosons.
In Fig.~\ref{fig} (right), $C_2^{W^\prime}/C_2^{\rm SM}$ is shown 
 on $g_{23}$--$g_{33}$ plane
 by fixing $M_V=1\,\TeV$
 and $g_{11}= - 3.6$ 
 corresponding to the maximum value allowed by the $K\to\pi\pi$ data in Eq.~\eqref{Kpipi}.
 Eventually, we obtain
\begin{align}
\frac{C_2^{W^\prime}(m_b)}{C_2^{\rm SM}(m_b) } \gtrsim - 0.10\,.
\end{align}

%%%%%%%%%%%%%%%%%%%%%%%%%%%%%%%%%%%%
\section{Discussion}
\label{sec:conclusion}
%%%%%%%%%%%%%%%%%%%%%%%%%%%%%%%%%%%%

Motivated by a recent improvement  of the SM predictions on
$\Bbar^0 \to D^{(\ast)+}K^-$ and 
$\Bbar^0_s \to D^{(\ast)+}_s \pi^-$, 
we investigated  the size of possible several new physics contributions to these processes. 
In spite of severe bounds from the other flavor observables and the LHC searches, 
we conclude that a $-10\%$ shift in the $b \to c \bar{u} q$ amplitude is possible by the  left-handed $W^\prime$ model.
Such a new physics contribution can reduce the tension
in the $b \to c \bar{u} q$ processes.

Since  $g_{22} = g_{11}$ is a necessary condition, 
this model also produces new physics contributions to 
$b \to c \bar{c} s$ processes with the same size \cite{Jager:2017gal,Jager:2019bgk}.
Although they, \eg, $B^+ \to J/\psi K^+$, have been measured precisely, 
the SM predictions suffer from large nonfactorizable corrections
\cite{Li:2006vq,Liu:2010zh,Li:2020app}.
We, therefore, expect that the $b \to c \bar{c} s$ processes are less sensitive than $b \to c \bar{u} q$.

It is unclear whether the new physics scalar operator can explain the $b\to c \bar{u}q$ anomaly, 
but
it is an interesting direction to consider it.
For instance, within a general two Higgs doublet model,
a charged Higgs interaction~is~\cite{Iguro:2017ysu}
\begin{align}
  {\cal L}= -H^+ \bar{u}^i(V\rho_d P_R-\rho_u^\dagger V P_L)_{ij} d^j  +{\rm H.c.}\,,
\end{align}
where $(V\rho_d)_{23}$ is stringently constrained by $\Delta M_s$ via a heavy neutral Higgs exchange, 
 while $(\rho_u^\dagger V)_{23}$ is less constrained
 by the flavor and collider observables   \cite{Iguro:2018qzf,Hou:2018zmg}.
 Therefore, a potentially large contribution to the $b\to c \bar{u}q$ processes would be expected.

%{\color{red}{
%please remove clearpage}}
%\clearpage

%%%%%%%%%%%%%%%%%%%%%%%%%%%%%%%%%%%%
%%%%%%%%%%%%%%%%%%%%%%%%%%%%%%%%%%%%
\section*{Acknowledgements}
%---------------------------------------------------------------------------
The authors thank  
Gauthier Durieux, Motoi Endo, Satoshi Mishima,  Yael Shadmi, Yotam Soreq
and 
Michihisa Takeuchi
for valuable comments and discussion on the analysis. 
%%%%%%%%%%%%%%%%%%%%%%%%%%%%%%%%%%%%
%%%%%%%%%%%%%%%
The work of S.I.\ is supported by the Japan Society for the Promotion of Science (JSPS) Research Fellowships for Young Scientists, No.19J10980.
S.I.\ would like to thank the warm hospitality at KEK where he stayed during the work.
%%%
%%%
The work of T.K.\ is supported in part by the JSPS Grant-in-Aid for 
Early-Career Scientists, No.19K14706.
%%%
The work is also supported by the JSPS Core-to-Core Program (Grant No.JPJSCCA20200002).

%---------------------------------------------------------------------------

\appendix

\begin{widetext}

%%%%%%%%%%%%%%%%%%%%%%%%%%%%%%%%%%%%

\section{Renormalization-group evolution}
\label{App:RG}

The LO RG evolution in the effective Lagrangian in Eq.~(\textcolor{blue}{5}) is 
given as \cite{Chetyrkin:1996vx}
\begin{align}
\frac{d \vec{C}(\mu)}{d \ln \mu} = \frac{\alpha_s(\mu)}{4 \pi} 
\begin{pmatrix}
-4 & 12 \\
\frac{8}{3} & 0
\end{pmatrix}\vec{C}(\mu)\,.
\end{align}
According to Ref.~\cite{Buras:1991jm},
we obtain an analytic solution of the LO RG evolution as 
\begin{align}
\begin{pmatrix}
C_1^{\rm NP} (m_W) \\
C_2^{\rm NP} (m_W)
\end{pmatrix}
= 
\begin{pmatrix}
 \frac{1}{3}\eta^{\frac{2}{7}} + \frac{2}{3}\eta^{- \frac{4}{7}}& 
 \eta^{\frac{2}{7}} -\eta^{- \frac{4}{7}}
\\
 \frac{2}{9}\eta^{\frac{2}{7}} - \frac{2}{9}\eta^{- \frac{4}{7}}  &  \frac{2}{3}\eta^{\frac{2}{7}} + \frac{1}{3}\eta^{- \frac{4}{7}}
\end{pmatrix} \begin{pmatrix}
C_1^{\rm NP} (\Lambda) \\
C_2^{\rm NP} (\Lambda)
\end{pmatrix}
\,,
\label{C2NP}
\end{align}
with $\eta = \alpha_s(\Lambda)/\alpha_s(m_W)$.
At the weak scale,
the SM contributions enter as \cite{Gorbahn:2004my}
\begin{align}
C_1^q(m_W) = 15  \frac{\alpha_s(m_W)}{4\pi}  + C_1^{q,{\rm NP}}(m_W)\,,
\qquad
C_2^q(m_W) = 1  + C_2^{q,{\rm NP}}(m_W)\,,
\end{align}
and their RG evaluation  from the weak scale to the hadronic scale is
\begin{align}
\begin{pmatrix}
C_1 (m_b) \\
C_2 (m_b)
\end{pmatrix}
= 
\begin{pmatrix}
 \frac{1}{3}\bar\eta^{\frac{6}{23}} + \frac{2}{3}\bar\eta^{- \frac{12}{23}}& 
 \bar\eta^{\frac{6}{23}} -\bar\eta^{- \frac{12}{23}}
\\
 \frac{2}{9}\bar\eta^{\frac{6}{23}} - \frac{2}{9}\bar\eta^{- \frac{12}{23}}  &  \frac{2}{3}\bar\eta^{\frac{6}{23}} + \frac{1}{3}\bar\eta^{- \frac{12}{23}}
\end{pmatrix} \begin{pmatrix}
C_1 (m_W) \\
C_2 (m_W)
\end{pmatrix}
\,,
\end{align}
with $\bar\eta = \alpha_s(m_W)/\alpha_s(m_b)$. 

%%%%%%%%%%%%%%%%%

%%%%%%%%%%%%%%%%%%%%%%%%%%%%%%%%%%%%
\section{Minimal flavor violation}
\label{sec:MFV}
%%%%%%%%%%%%%%%%%%%%%%%%%%%%%%%%%%%%

In this section, we give the detailed calculations 
for the MFV scenario.
In the MFV hypothesis, 
the SU(3$)_{Q_L} \times $SU(3$)_{U_R} \times $SU(3$)_{D_R} $ flavor symmetry is introduced and it is broken  
only by the Yukawa interactions \cite{DAmbrosio:2002vsn,Isidori:2012ts}.
Under this hypothesis, the flavor structure is the same as the SM one: 
the flavor-changing neutral currents are automatically suppressed.
For the $b\to c\bar{u}q$ anomaly,
we consider the following dimension-six operator,
\begin{align}
\mathcal{L} = \frac{1}{2 \Lambda^2} \left\{ \bar{Q}_L^i  \left[\delta_{ij} + a ( Y^u Y^{u \dag})_{i\neq j} \right]\gamma^{\mu}  Q_L^j \right\}^2\,,
\label{eq:MFVOP}
\end{align}
with $Y^u = V^{\dag} \textrm{diag}(y_u, y_c, y_t)$, and $a$ is a dimensionless coupling.
In the quark mass-diagonal basis ($u_L^{\rm diag} = V u_L,~d_L^{\rm diag} =  d_L$), 
this operator produces
\begin{align}
\mathcal{L} \simeq \frac{1}{\Lambda^2} (V_{cb} + a y_t^2 
%V_{cs} V_{tb} \sim 1
V_{ts}^{\ast} ) V_{uq}^{\ast} \left(\bar{c}_L \gamma^{\mu } b_L \right)
\left(\bar{q}_L \gamma_{\mu} u_L \right)\,.
\end{align}
So, we obtain
\begin{align}
C_{2}^{q,{\rm MFV}} (\Lambda)= - \frac{1}{\Lambda^2}
\frac{\sqrt{2}}{4 G_F} \left(1 + a y_t^2 \frac{
%V_{cs} V_{tb}\sim 1
V_{ts}^{\ast}}{V_{cb}} \right)\,.
\end{align}

From the operator in Eq.~\eqref{eq:MFVOP}, 
we also obtain a constraint from 
the $B_{s}$-meson mass difference ($\Delta M_{{s}}$)
 as (cf., Ref.~\cite{Silvestrini:2018dos}), 
 %(cf., Ref.~\cite{Silvestrini:2018dos} provides a bound of 6.2\,\TeV), 
\begin{align}
\left| \Lambda/a \right| \gtrsim 7.9\,\TeV\,, 
\end{align}
where the LO RG effect is taken into account \cite{Bagger:1997gg}, 
\begin{align}
C_{LL} (m_W) = \eta^{\frac{2}{7}} C_{LL} (\Lambda)\,,
\end{align}  
with $\eta = \alpha_s(\Lambda)/\alpha_s(m_W)$, 
  and the latest SM estimation of $\Delta M_s$
 is adopted \cite{Blanke:2018cya,DiLuzio:2019jyq}. 
We required the new physics contribution to $\Delta M_s$ does not change the SM prediction at $2\sigma$ level.

On the other hand, 
from the $b\to c \bar{u} q$ anomaly in Eq.~(\textcolor{blue}{9}), 
we find
\begin{align}
\Lambda \sim \sqrt{1 - a} \left( 0.43 {}^{+ 5}_{-3} \right) \TeV\,.
\end{align}
Therefore, 
we reach a requirement for the anomaly:
\begin{align}
\Lambda \lesssim 0.49\,\TeV \,\textrm{~and~}\, |a| \lesssim 0.06 \,.
\label{eq:anomalyWC}
\end{align}

However, such a contact interaction can be probed by 
a non-resonant dijet angular distribution search in the LHC.
The result is reported by the ATLAS collaboration 
at $\sqrt{s} = 13\,\TeV$ with $\int dt \mathcal{L}=37$\,fb$^{-1}$ \cite{Aaboud:2017yvp}.
We interpret the result in terms of  the operator in Eq.~\eqref{eq:MFVOP}, and obtain a $95\%$ CL exclusion limit,
\begin{align}
\Lambda < 3.7\,\TeV \,\textrm{~and~}\,  4.9\,\TeV < \Lambda < 8.3\,\TeV\,.
\end{align}
This bound is clearly incompatible with Eq.~\eqref{eq:anomalyWC}.
From this constraint, we obtain a bound
\begin{align}
\frac{C_2^{\rm MFV}(m_b)}{C_2^{\rm SM}(m_b)}  \gtrsim -0.002\,.
\end{align}
Therefore, this new physics scenario never explains the $b\to c \bar{u}q$ anomaly.

%%%%%%%%%%%%%%%%%%%%%%%

\section{Loop functions}
\label{App:loop}

 The loop functions $f(x)$ and $f^\prime(x,y)$ in Eq.~(\textcolor{blue}{20}) are defined by \cite{Lee:1998qq} (cf.~\cite{Choudhury:2004bh})
\begin{align}
f(x) =& \frac{4  - 11 x + x^2}{4 (1-x)^2} - \frac{3 x^2 {\rm{ln}}\,x}{2 (1-x)^3}\,,\\
f^\prime (x,y) =& \frac{1}{4 y (x-y)^2 (1-x)^2}  \biggl[(1-x) (4 x^2 + 4 y^2 + 5 x^2 y - 8 x y - 4 x y^2 - x^3)\nonumber\\
&- 3 x^2 (x - 2 y + x y ) {\rm{ln}} \left( \frac{ y}{x}\right) - \frac{3 x (x-y)^2}{y-1} {\rm{ln}}\, y  \biggr]\,,
\label{f2p}
\end{align}
where 
$\lim_{y\to 1} f^\prime (x,y) = f (x)$.
We note that Ref.~\cite{Lee:1998qq} contains a typo in its Eq.~(22),
where $-x^{2}$ in the last term of the first line in the arXiv version must be replaced by $-x^{3}$.

The loop functions $f_{\gamma}(x)$ and $f_{g}(x)$ in Eq.~\eqref{eq:C7C8} are defined by \cite{Grinstein:1990tj}
\begin{align}
f_{\gamma}(x) & = \frac{3 x^3 - 2 x^2}{4 (x-1)^4} \ln x + \frac{- 8 x^3 - 5 x^2 + 7 x}{24 (x-1)^3}\,,\\
f_{g}(x) & = \frac{- 3 x^2 }{4 (x-1)^4} \ln x + \frac{-  x^3 + 5 x^2 + 2 x}{8 (x-1)^3}\,.
\end{align}

\section{\boldmath{$W^\prime$} and \boldmath{$Z^\prime$} contributions to \boldmath{$b \to s \gamma$}}
\label{App:bsg}

The effective Lagrangian for the $b \to s \gamma$ process is
\begin{align}
\mathcal{L}  = \frac{G_F}{\sqrt{2}} V_{ts}^{\ast} V_{tb} \left[ \sum_{i=1}^6 C_{i} (\mu) \mathcal{Q}_{i} (\mu)
+
C_{7 \gamma} (\mu) Q_{7 \gamma} (\mu)
+ 
 C_{8g} (\mu) Q_{8 g} (\mu)\right]\,,
\end{align}
with
\begin{align}
Q_{7\gamma} = \frac{e}{8 \pi^2} m_b \bar{s} \sigma^{\mu \nu} (1+\gamma_5) b F_{\mu \nu}\,,
\qquad
Q_{8 g } = \frac{g_s}{8 \pi^2} m_b \bar{s}\sigma^{\mu \nu}T^a (1+\gamma_5) b G^a_{\mu \nu}\,,
\end{align}
and the operators $\mathcal{Q}_{1}$--$\mathcal{Q}_{6}$ are defined in Ref.~\cite{Buras:1998raa}.

By integrating out the heavy gauge bosons, we obtain
\begin{alignat}{2}
C_2 (M_V) &\simeq  \frac{g_{11}g_{33}}{g_W^2} \frac{m_W^2}{M_V^2}\,,
 &
C_3 (M_V) &\simeq - \frac{g_{11}g_{23}}{2 g_W^2 V_{ts}^{\ast}} \frac{m_W^2}{M_V^2}\,,\\
C_7 (M_V) &=\frac{g_{11} g_{33}}{g_W^2} \frac{m_W^2}{M_V^2} f_{\gamma}\left( \frac{m_t^2}{M_V^2}\right)  \,,
\qquad&
C_8 (M_V) &=\frac{g_{11} g_{33}}{g_W^2} \frac{m_W^2}{M_V^2} f_{g}\left( \frac{m_t^2}{M_V^2}\right)   \,,
\label{eq:C7C8}
\end{alignat}
and remaining coefficients are set to zero at $\mu = M_V$.
To obtain new physics contributions at the hadronic scale, we solved the corresponding RG evolution down to $\mu = m_b$ numerically:
\begin{align}
\frac{d \vec{C}(\mu)}{d \ln \mu} = \frac{\alpha_s (\mu)}{4 \pi} \left(\hat{\gamma}^{(0){\rm eff}}\right)^T \vec{C}(\mu)\,, 
\qquad
\vec{C} = 
\left(C_1,\,C_2,\, \cdots,\, C_6,\, C_7^{\rm eff},\, C_8^{\rm eff} \right)\,,
\end{align}
where the anomalous dimension matrix $\hat{\gamma}^{(0){\rm eff}}$ is given in Refs.~\cite{Ciuchini:1993fk,Buras:1998raa}.
The $ C_7^{\rm eff},\, C_8^{\rm eff}$ are the effective Wilson coefficients which are required to cancel  a regularization scheme dependence \cite{Buras:1993xp}.
In this model, $C_7^{\rm eff} (M_V) = C_7 (M_V)$ and  $C_8^{\rm eff} (M_V) = C_8 (M_V)$.

Using $C_7^{\rm eff}(m_b)$, we obtain a constraint from $b \to s \gamma$ data, where  we required the new physics contributions are within a $2\sigma$ uncertainty range
\cite{Misiak:2015xwa,Endo:2017ums}.
The bound is sensitive to $g_{23}$ which comes from the $Z^\prime$ contribution to $C_{3}(M_V)$.
 For $g_{23} =0$, we obtain
 \begin{align}
\sqrt{|g_{11}g_{33}|}/M_V  \lesssim 2.5 \,\left(\TeV\right)^{-1}\,.
 \end{align}
 This bound is significantly alleviated for $g_{23}/g_{33} >0$ region, while it becomes  stronger for $g_{23}/g_{33} <0$ region.

\end{widetext}

%%%%%%%%%%%%%%%%%%%%%%%%%%%%%%%%%%%%
%%%%%%%%%%%%%%%%%%%%%%%%%%%%%%%%%%%%

\bibliography{ref}

\providecommand{\href}[2]{#2}\begingroup\raggedright\begin{thebibliography}{10}

\bibitem{Bordone:2020gao}
M.~Bordone, N.~Gubernari, T.~Huber, M.~Jung, and D.~van Dyk, {\em {A puzzle in
  $\bar{B}_{(s)}^0 \to D_{(s)}^+ \lbrace \pi^-, K^-\rbrace$ decays and
  extraction of the $f_s/f_d$ fragmentation fraction}.} {\ttfamily
  \href{https://arxiv.org/abs/2007.10338}{arXiv:2007.10338}}.

\bibitem{PDG2020}
{\bfseries Particle Data Group} Collaboration, {\em {Review of Particle
  Physics},} \href{https://dx.doi.org/10.1093/ptep/ptaa104}{Prog.\  Theor.\
  Exp.\  Phys.\  {\bfseries 2020} (2020) 083C01}.

\bibitem{Beneke:1999br}
M.~Beneke, G.~Buchalla, M.~Neubert, and C.~T.~Sachrajda, {\em {QCD
  factorization for $B \to \pi \pi$ decays: Strong phases and CP violation in
  the heavy quark limit},}
  \href{https://dx.doi.org/10.1103/PhysRevLett.83.1914}{Phys.\  Rev.\  Lett.\
  {\bfseries 83} (1999) 1914--1917} {\ttfamily
  [\href{https://arxiv.org/abs/hep-ph/9905312}{hep-ph/9905312}]}.

\bibitem{Beneke:2000ry}
M.~Beneke, G.~Buchalla, M.~Neubert, and C.~T.~Sachrajda, {\em {QCD
  factorization for exclusive, nonleptonic B meson decays: General arguments
  and the case of heavy light final states},}
  \href{https://dx.doi.org/10.1016/S0550-3213(00)00559-9}{Nucl.\  Phys.\  B
  {\bfseries 591} (2000) 313--418} {\ttfamily
  [\href{https://arxiv.org/abs/hep-ph/0006124}{hep-ph/0006124}]}.

\bibitem{Beneke:2001ev}
M.~Beneke, G.~Buchalla, M.~Neubert, and C.~T.~Sachrajda, {\em {QCD
  factorization in $B \to \pi K, \pi \pi$ decays and extraction of Wolfenstein
  parameters},} \href{https://dx.doi.org/10.1016/S0550-3213(01)00251-6}{Nucl.\
  Phys.\  B {\bfseries 606} (2001) 245--321} {\ttfamily
  [\href{https://arxiv.org/abs/hep-ph/0104110}{hep-ph/0104110}]}.

\bibitem{Gorbahn:2004my}
M.~Gorbahn and U.~Haisch, {\em {Effective Hamiltonian for non-leptonic $|\Delta
  F| = 1$ decays at NNLO in QCD},}
  \href{https://dx.doi.org/10.1016/j.nuclphysb.2005.01.047}{Nucl.\  Phys.\  B
  {\bfseries 713} (2005) 291--332} {\ttfamily
  [\href{https://arxiv.org/abs/hep-ph/0411071}{hep-ph/0411071}]}.

\bibitem{Huber:2016xod}
T.~Huber, S.~Kränkl, and X.-Q.~Li, {\em {Two-body non-leptonic heavy-to-heavy
  decays at NNLO in QCD factorization},}
  \href{https://dx.doi.org/10.1007/JHEP09(2016)112}{JHEP {\bfseries 09} (2016)
  112} {\ttfamily [\href{https://arxiv.org/abs/1606.02888}{arXiv:1606.02888}]}.

\bibitem{Jung:2018lfu}
M.~Jung and D.~M.~Straub, {\em {Constraining new physics in $b\to c\ell\nu$
  transitions},} \href{https://dx.doi.org/10.1007/JHEP01(2019)009}{JHEP
  {\bfseries 01} (2019) 009} {\ttfamily
  [\href{https://arxiv.org/abs/1801.01112}{arXiv:1801.01112}]}.

\bibitem{Bordone:2019vic}
M.~Bordone, M.~Jung, and D.~van Dyk, {\em {Theory determination of $\bar{B}\to
  D^{(*)}\ell^-\bar\nu$ form factors at $\mathcal{O}(1/m_c^2)$},}
  \href{https://dx.doi.org/10.1140/epjc/s10052-020-7616-4}{Eur.\  Phys.\  J.\
  C {\bfseries 80} (2020) 74} {\ttfamily
  [\href{https://arxiv.org/abs/1908.09398}{arXiv:1908.09398}]}.

\bibitem{Bordone:2019guc}
M.~Bordone, N.~Gubernari, D.~van Dyk, and M.~Jung, {\em {Heavy-Quark expansion
  for ${{\bar{B}}_s\rightarrow D^{(*)}_s}$ form factors and unitarity bounds
  beyond the ${SU(3)_F}$ limit},}
  \href{https://dx.doi.org/10.1140/epjc/s10052-020-7850-9}{Eur.\  Phys.\  J.\
  C {\bfseries 80} (2020) 347} {\ttfamily
  [\href{https://arxiv.org/abs/1912.09335}{arXiv:1912.09335}]}.

\bibitem{Iguro:2020cpg}
S.~Iguro and R.~Watanabe, {\em {Bayesian fit analysis to full distribution data
  of $\bar B \to D^{(*)} \ell\bar\nu$: $|V_{cb}|$ determination and New Physics
  constraints},} \href{https://dx.doi.org/10.1007/JHEP08(2020)006}{JHEP
  {\bfseries 08} (2020) 006} {\ttfamily
  [\href{https://arxiv.org/abs/2004.10208}{arXiv:2004.10208}]}.

\bibitem{Amhis:2019ckw}
{\bfseries HFLAV} Collaboration, {\em {Averages of $b$-hadron, $c$-hadron, and
  $\tau$-lepton properties as of 2018}.} {\ttfamily
  \href{https://arxiv.org/abs/1909.12524}{arXiv:1909.12524}}. 2019 updated
  results and plots available at \url{https://hflav.web.cern.ch/}.

\bibitem{Kim:2019vic}
{\bfseries LANL-SWME} Collaboration, {\em {2019 Update on $\varepsilon_K$ with
  lattice QCD inputs},} \href{https://dx.doi.org/10.22323/1.363.0029}{PoS
  {\bfseries LATTICE2019} (2019) 029} {\ttfamily
  [\href{https://arxiv.org/abs/1912.03024}{arXiv:1912.03024}]}.

\bibitem{Chetyrkin:1996vx}
K.~G.~Chetyrkin, M.~Misiak, and M.~Munz, {\em {Weak radiative B meson decay
  beyond leading logarithms},}
  \href{https://dx.doi.org/10.1016/S0370-2693(97)00324-9}{Phys.\  Lett.\  B
  {\bfseries 400} (1997) 206--219} {\ttfamily
  [\href{https://arxiv.org/abs/hep-ph/9612313}{hep-ph/9612313}]}. [Erratum:
  Phys.Lett.B 425, 414 (1998)].

\bibitem{Chetyrkin:1997gb}
K.~G.~Chetyrkin, M.~Misiak, and M.~Munz, {\em {$|\Delta F| = 1$ nonleptonic
  effective Hamiltonian in a simpler scheme},}
  \href{https://dx.doi.org/10.1016/S0550-3213(98)00131-X}{Nucl.\  Phys.\  B
  {\bfseries 520} (1998) 279--297} {\ttfamily
  [\href{https://arxiv.org/abs/hep-ph/9711280}{hep-ph/9711280}]}.

\bibitem{Cabibbo:1963yz}
N.~Cabibbo, {\em {Unitary Symmetry and Leptonic Decays},}
  \href{https://dx.doi.org/10.1103/PhysRevLett.10.531}{Phys.\  Rev.\  Lett.\
  {\bfseries 10} (1963) 531--533}.

\bibitem{Kobayashi:1973fv}
M.~Kobayashi and T.~Maskawa, {\em {CP Violation in the Renormalizable Theory of
  Weak Interaction},} \href{https://dx.doi.org/10.1143/PTP.49.652}{Prog.\
  Theor.\  Phys.\  {\bfseries 49} (1973) 652--657}.

\bibitem{Bobeth:2014rda}
C.~Bobeth, U.~Haisch, A.~Lenz, B.~Pecjak, and G.~Tetlalmatzi-Xolocotzi, {\em
  {On new physics in $\Delta\Gamma_{d}$},}
  \href{https://dx.doi.org/10.1007/JHEP06(2014)040}{JHEP {\bfseries 06} (2014)
  040} {\ttfamily [\href{https://arxiv.org/abs/1404.2531}{arXiv:1404.2531}]}.

\bibitem{Brod:2014bfa}
J.~Brod, A.~Lenz, G.~Tetlalmatzi-Xolocotzi, and M.~Wiebusch, {\em {New physics
  effects in tree-level decays and the precision in the determination of the
  quark mixing angle $\gamma$},}
  \href{https://dx.doi.org/10.1103/PhysRevD.92.033002}{Phys.\  Rev.\  D
  {\bfseries 92} (2015) 033002} {\ttfamily
  [\href{https://arxiv.org/abs/1412.1446}{arXiv:1412.1446}]}.

\bibitem{Lenz:2019lvd}
A.~Lenz and G.~Tetlalmatzi-Xolocotzi, {\em {Model-independent bounds on new
  physics effects in non-leptonic tree-level decays of B-mesons},}
  \href{https://dx.doi.org/10.1007/JHEP07(2020)177}{JHEP {\bfseries 07} (2020)
  177} {\ttfamily [\href{https://arxiv.org/abs/1912.07621}{arXiv:1912.07621}]}.

\bibitem{Kitahara:2016nld}
T.~Kitahara, U.~Nierste, and P.~Tremper, {\em {Singularity-free next-to-leading
  order $\Delta$S = 1 renormalization group evolution and
  $\epsilon_K'/\epsilon_K$ in the Standard Model and beyond},}
  \href{https://dx.doi.org/10.1007/JHEP12(2016)078}{JHEP {\bfseries 12} (2016)
  078} {\ttfamily [\href{https://arxiv.org/abs/1607.06727}{arXiv:1607.06727}]}.

\bibitem{Blum:2015ywa}
T.~Blum {\em et~al.}, {\em {$K \rightarrow \pi\pi$ $\Delta I=3/2$ decay
  amplitude in the continuum limit},}
  \href{https://dx.doi.org/10.1103/PhysRevD.91.074502}{Phys.\  Rev.\  D
  {\bfseries 91} (2015) 074502} {\ttfamily
  [\href{https://arxiv.org/abs/1502.00263}{arXiv:1502.00263}]}.

\bibitem{Blum:2011ng}
T.~Blum {\em et~al.}, {\em {The $K\to(\pi\pi)_{I=2}$ Decay Amplitude from
  Lattice QCD},}
  \href{https://dx.doi.org/10.1103/PhysRevLett.108.141601}{Phys.\  Rev.\
  Lett.\  {\bfseries 108} (2012) 141601} {\ttfamily
  [\href{https://arxiv.org/abs/1111.1699}{arXiv:1111.1699}]}.

\bibitem{Bai:2015nea}
{\bfseries RBC, UKQCD} Collaboration, {\em {Standard Model Prediction for
  Direct CP Violation in K$\rightarrow$$\pi$$\pi$ Decay},}
  \href{https://dx.doi.org/10.1103/PhysRevLett.115.212001}{Phys.\  Rev.\
  Lett.\  {\bfseries 115} (2015) 212001} {\ttfamily
  [\href{https://arxiv.org/abs/1505.07863}{arXiv:1505.07863}]}.

\bibitem{Ishizuka:2018qbn}
N.~Ishizuka, K.~I.~Ishikawa, A.~Ukawa, and T.~Yoshié, {\em {Calculation of $K
  \to \pi\pi$ decay amplitudes with improved Wilson fermion action in non-zero
  momentum frame in lattice QCD},}
  \href{https://dx.doi.org/10.1103/PhysRevD.98.114512}{Phys.\  Rev.\  D
  {\bfseries 98} (2018) 114512} {\ttfamily
  [\href{https://arxiv.org/abs/1809.03893}{arXiv:1809.03893}]}.

\bibitem{Abbott:2020hxn}
{\bfseries RBC, UKQCD} Collaboration, {\em {Direct CP violation and the $\Delta
  I=1/2$ rule in $K\to\pi\pi$ decay from the standard model},}
  \href{https://dx.doi.org/10.1103/PhysRevD.102.054509}{Phys.\  Rev.\  D
  {\bfseries 102} (2020) 054509} {\ttfamily
  [\href{https://arxiv.org/abs/2004.09440}{arXiv:2004.09440}]}.

\bibitem{DAmbrosio:2002vsn}
G.~D'Ambrosio, G.~F.~Giudice, G.~Isidori, and A.~Strumia, {\em {Minimal flavor
  violation: An Effective field theory approach},}
  \href{https://dx.doi.org/10.1016/S0550-3213(02)00836-2}{Nucl.\  Phys.\  B
  {\bfseries 645} (2002) 155--187} {\ttfamily
  [\href{https://arxiv.org/abs/hep-ph/0207036}{hep-ph/0207036}]}.

\bibitem{Isidori:2012ts}
G.~Isidori and D.~M.~Straub, {\em {Minimal Flavour Violation and Beyond},}
  \href{https://dx.doi.org/10.1140/epjc/s10052-012-2103-1}{Eur.\  Phys.\  J.\
  C {\bfseries 72} (2012) 2103} {\ttfamily
  [\href{https://arxiv.org/abs/1202.0464}{arXiv:1202.0464}]}.

\bibitem{Aaboud:2017yvp}
{\bfseries ATLAS} Collaboration, {\em {Search for new phenomena in dijet events
  using 37 fb$^{-1}$ of $pp$ collision data collected at $\sqrt{s}=$13 TeV with
  the ATLAS detector},}
  \href{https://dx.doi.org/10.1103/PhysRevD.96.052004}{Phys.\  Rev.\  D
  {\bfseries 96} (2017) 052004} {\ttfamily
  [\href{https://arxiv.org/abs/1703.09127}{arXiv:1703.09127}]}.

\bibitem{Langacker:1988ur}
P.~Langacker and D.~London, {\em {Mixing Between Ordinary and Exotic
  Fermions},} \href{https://dx.doi.org/10.1103/PhysRevD.38.886}{Phys.\  Rev.\
  D {\bfseries 38} (1988) 886}.

\bibitem{Chivukula:2003wj}
R.~S.~Chivukula, H.-J.~He, J.~Howard, and E.~H.~Simmons, {\em {The Structure of
  electroweak corrections due to extended gauge symmetries},}
  \href{https://dx.doi.org/10.1103/PhysRevD.69.015009}{Phys.\  Rev.\  D
  {\bfseries 69} (2004) 015009} {\ttfamily
  [\href{https://arxiv.org/abs/hep-ph/0307209}{hep-ph/0307209}]}.

\bibitem{Chiang:2009kb}
C.-W.~Chiang, N.~G.~Deshpande, X.-G.~He, and J.~Jiang, {\em {The Family
  SU(2$)_l \times$ SU(2$)_h \times$ U(1) Model},}
  \href{https://dx.doi.org/10.1103/PhysRevD.81.015006}{Phys.\  Rev.\  D
  {\bfseries 81} (2010) 015006} {\ttfamily
  [\href{https://arxiv.org/abs/0911.1480}{arXiv:0911.1480}]}.

\bibitem{Boucenna:2016wpr}
S.~M.~Boucenna, A.~Celis, J.~Fuentes-Martin, A.~Vicente, and J.~Virto, {\em
  {Non-abelian gauge extensions for B-decay anomalies},}
  \href{https://dx.doi.org/10.1016/j.physletb.2016.06.067}{Phys.\  Lett.\  B
  {\bfseries 760} (2016) 214--219} {\ttfamily
  [\href{https://arxiv.org/abs/1604.03088}{arXiv:1604.03088}]}.

\bibitem{Boucenna:2016qad}
S.~M.~Boucenna, A.~Celis, J.~Fuentes-Martin, A.~Vicente, and J.~Virto, {\em
  {Phenomenology of an $SU(2) \times SU(2) \times U(1)$ model with
  lepton-flavour non-universality},}
  \href{https://dx.doi.org/10.1007/JHEP12(2016)059}{JHEP {\bfseries 12} (2016)
  059} {\ttfamily [\href{https://arxiv.org/abs/1608.01349}{arXiv:1608.01349}]}.

\bibitem{Golowich:2007ka}
E.~Golowich, J.~Hewett, S.~Pakvasa, and A.~A.~Petrov, {\em {Implications of
  $D^0$ - $\bar{D}^0$ Mixing for New Physics},}
  \href{https://dx.doi.org/10.1103/PhysRevD.76.095009}{Phys.\  Rev.\  D
  {\bfseries 76} (2007) 095009} {\ttfamily
  [\href{https://arxiv.org/abs/0705.3650}{arXiv:0705.3650}]}.

\bibitem{Barbieri:1995uv}
R.~Barbieri, G.~R.~Dvali, and L.~J.~Hall, {\em {Predictions from a U(2) flavor
  symmetry in supersymmetric theories},}
  \href{https://dx.doi.org/10.1016/0370-2693(96)00318-8}{Phys.\  Lett.\  B
  {\bfseries 377} (1996) 76--82} {\ttfamily
  [\href{https://arxiv.org/abs/hep-ph/9512388}{hep-ph/9512388}]}.

\bibitem{Barbieri:2011ci}
R.~Barbieri, G.~Isidori, J.~Jones-Perez, P.~Lodone, and D.~M.~Straub, {\em
  {$U(2)$ and Minimal Flavour Violation in Supersymmetry},}
  \href{https://dx.doi.org/10.1140/epjc/s10052-011-1725-z}{Eur.\  Phys.\  J.\
  C {\bfseries 71} (2011) 1725} {\ttfamily
  [\href{https://arxiv.org/abs/1105.2296}{arXiv:1105.2296}]}.

\bibitem{Blanke:2018cya}
M.~Blanke and A.~J.~Buras, {\em {Emerging $\Delta M_{d}$ -anomaly from
  tree-level determinations of $|V_{cb}|$ and the angle $\gamma $},}
  \href{https://dx.doi.org/10.1140/epjc/s10052-019-6667-x}{Eur.\  Phys.\  J.\
  C {\bfseries 79} (2019) 159} {\ttfamily
  [\href{https://arxiv.org/abs/1812.06963}{arXiv:1812.06963}]}.

\bibitem{DiLuzio:2019jyq}
L.~Di~Luzio, M.~Kirk, A.~Lenz, and T.~Rauh, {\em {$\Delta M_s$ theory precision
  confronts flavour anomalies},}
  \href{https://dx.doi.org/10.1007/JHEP12(2019)009}{JHEP {\bfseries 12} (2019)
  009} {\ttfamily [\href{https://arxiv.org/abs/1909.11087}{arXiv:1909.11087}]}.

\bibitem{Aad:2019hjw}
{\bfseries ATLAS} Collaboration, {\em {Search for new resonances in mass
  distributions of jet pairs using 139 fb$^{-1}$ of $pp$ collisions at
  $\sqrt{s}=13$ TeV with the ATLAS detector},}
  \href{https://dx.doi.org/10.1007/JHEP03(2020)145}{JHEP {\bfseries 03} (2020)
  145} {\ttfamily [\href{https://arxiv.org/abs/1910.08447}{arXiv:1910.08447}]}.

\bibitem{Sirunyan:2019vgj}
{\bfseries CMS} Collaboration, {\em {Search for high mass dijet resonances with
  a new background prediction method in proton-proton collisions at $\sqrt{s}
  =$ 13 TeV},} \href{https://dx.doi.org/10.1007/JHEP05(2020)033}{JHEP
  {\bfseries 05} (2020) 033} {\ttfamily
  [\href{https://arxiv.org/abs/1911.03947}{arXiv:1911.03947}]}.

\bibitem{Sirunyan:2018xlo}
{\bfseries CMS} Collaboration, {\em {Search for narrow and broad dijet
  resonances in proton-proton collisions at $ \sqrt{s}=13 $ TeV and constraints
  on dark matter mediators and other new particles},}
  \href{https://dx.doi.org/10.1007/JHEP08(2018)130}{JHEP {\bfseries 08} (2018)
  130} {\ttfamily [\href{https://arxiv.org/abs/1806.00843}{arXiv:1806.00843}]}.

\bibitem{Sirunyan:2018ryr}
{\bfseries CMS} Collaboration, {\em {Search for resonant $
  \mathrm{t}\overline{\mathrm{t}} $ production in proton-proton collisions at $
  \sqrt{s}=13 $ TeV},} \href{https://dx.doi.org/10.1007/JHEP04(2019)031}{JHEP
  {\bfseries 04} (2019) 031} {\ttfamily
  [\href{https://arxiv.org/abs/1810.05905}{arXiv:1810.05905}]}.

\bibitem{Aad:2020kop}
{\bfseries ATLAS} Collaboration, {\em {Search for $t\bar{t}$ resonances in
  fully hadronic final states in $pp$ collisions at $\sqrt{s}=13$ TeV with the
  ATLAS detector}.} {\ttfamily
  \href{https://arxiv.org/abs/2005.05138}{arXiv:2005.05138}}.

\bibitem{Sirunyan:2017vkm}
{\bfseries CMS} Collaboration, {\em {Search for heavy resonances decaying to a
  top quark and a bottom quark in the lepton+jets final state in proton--proton
  collisions at 13 TeV},}
  \href{https://dx.doi.org/10.1016/j.physletb.2017.12.006}{Phys.\  Lett.\  B
  {\bfseries 777} (2018) 39--63} {\ttfamily
  [\href{https://arxiv.org/abs/1708.08539}{arXiv:1708.08539}]}.

\bibitem{Aaboud:2018jux}
{\bfseries ATLAS} Collaboration, {\em {Search for vector-boson resonances
  decaying to a top quark and bottom quark in the lepton plus jets final state
  in $pp$ collisions at $\sqrt{s}$ = 13 TeV with the ATLAS detector},}
  \href{https://dx.doi.org/10.1016/j.physletb.2018.11.032}{Phys.\  Lett.\  B
  {\bfseries 788} (2019) 347--370} {\ttfamily
  [\href{https://arxiv.org/abs/1807.10473}{arXiv:1807.10473}]}.

\bibitem{Hayreter:2019dzc}
A.~Hayreter, X.-G.~He, and G.~Valencia, {\em {LHC constraints on
  $W^\prime,~Z^\prime$ that couple mainly to third generation fermions}.}
  {\ttfamily \href{https://arxiv.org/abs/1912.06344}{arXiv:1912.06344}}.

\bibitem{Abe:2015uaa}
T.~Abe, T.~Kitahara, and M.~M.~Nojiri, {\em {Prospects for Spin-1 Resonance
  Search at 13 TeV LHC and the ATLAS Diboson Excess},}
  \href{https://dx.doi.org/10.1007/JHEP02(2016)084}{JHEP {\bfseries 02} (2016)
  084} {\ttfamily [\href{https://arxiv.org/abs/1507.01681}{arXiv:1507.01681}]}.

\bibitem{ATLAS:2012pu}
{\bfseries ATLAS} Collaboration, {\em {ATLAS search for new phenomena in dijet
  mass and angular distributions using $pp$ collisions at $\sqrt{s}=7$ TeV},}
  \href{https://dx.doi.org/10.1007/JHEP01(2013)029}{JHEP {\bfseries 01} (2013)
  029} {\ttfamily [\href{https://arxiv.org/abs/1210.1718}{arXiv:1210.1718}]}.

\bibitem{Jager:2017gal}
S.~Jäger, M.~Kirk, A.~Lenz, and K.~Leslie, {\em {Charming new physics in rare
  B-decays and mixing?}}
  \href{https://dx.doi.org/10.1103/PhysRevD.97.015021}{Phys.\  Rev.\  D
  {\bfseries 97} (2018) 015021} {\ttfamily
  [\href{https://arxiv.org/abs/1701.09183}{arXiv:1701.09183}]}.

\bibitem{Jager:2019bgk}
S.~Jäger, M.~Kirk, A.~Lenz, and K.~Leslie, {\em {Charming New $B$-Physics},}
  \href{https://dx.doi.org/10.1007/JHEP03(2020)122}{JHEP {\bfseries 03} (2020)
  122} {\ttfamily [\href{https://arxiv.org/abs/1910.12924}{arXiv:1910.12924}]}.

\bibitem{Li:2006vq}
H.-n.~Li and S.~Mishima, {\em {Penguin pollution in the $B^0 \to J/\psi K_S$
  decay},} \href{https://dx.doi.org/10.1088/1126-6708/2007/03/009}{JHEP
  {\bfseries 03} (2007) 009} {\ttfamily
  [\href{https://arxiv.org/abs/hep-ph/0610120}{hep-ph/0610120}]}.

\bibitem{Liu:2010zh}
X.~Liu, Z.-Q.~Zhang, and Z.-J.~Xiao, {\em {$B \to (J/\Psi, \eta_c) K$ decays in
  the perturbative QCD approach},}
  \href{https://dx.doi.org/10.1088/1674-1137/34/7/002}{Chin.\  Phys.\  C
  {\bfseries 34} (2010) 937--943} {\ttfamily
  [\href{https://arxiv.org/abs/0901.0165}{arXiv:0901.0165}]}.

\bibitem{Li:2020app}
Y.-Q.~Li, M.-K.~Jia, and Z.~Rui, {\em {Revisiting nonfactorizable contributions
  to factorization-forbidden decays of $B$ mesons to charmonium},}
  \href{https://dx.doi.org/10.1088/1674-1137/abae50}{Chin.\  Phys.\  C
  {\bfseries 44} (2020) 113104} {\ttfamily
  [\href{https://arxiv.org/abs/2006.07625}{arXiv:2006.07625}]}.

\bibitem{Iguro:2017ysu}
S.~Iguro and K.~Tobe, {\em {$R(D^{(*)})$ in a general two Higgs doublet
  model},} \href{https://dx.doi.org/10.1016/j.nuclphysb.2017.10.014}{Nucl.\
  Phys.\  B {\bfseries 925} (2017) 560--606} {\ttfamily
  [\href{https://arxiv.org/abs/1708.06176}{arXiv:1708.06176}]}.

\bibitem{Iguro:2018qzf}
S.~Iguro and Y.~Omura, {\em {Status of the semileptonic $B$ decays and muon g-2
  in general 2HDMs with right-handed neutrinos},}
  \href{https://dx.doi.org/10.1007/JHEP05(2018)173}{JHEP {\bfseries 05} (2018)
  173} {\ttfamily [\href{https://arxiv.org/abs/1802.01732}{arXiv:1802.01732}]}.

\bibitem{Hou:2018zmg}
W.-S.~Hou, M.~Kohda, and T.~Modak, {\em {Constraining a Lighter Exotic Scalar
  via Same-sign Top},}
  \href{https://dx.doi.org/10.1016/j.physletb.2018.09.046}{Phys.\  Lett.\  B
  {\bfseries 786} (2018) 212--216} {\ttfamily
  [\href{https://arxiv.org/abs/1808.00333}{arXiv:1808.00333}]}.

\bibitem{Buras:1991jm}
A.~J.~Buras, M.~Jamin, M.~E.~Lautenbacher, and P.~H.~Weisz, {\em {Effective
  Hamiltonians for $\Delta S = 1$ and $\Delta B = 1$ nonleptonic decays beyond
  the leading logarithmic approximation},}
  \href{https://dx.doi.org/10.1016/0550-3213(92)90345-C}{Nucl.\  Phys.\  B
  {\bfseries 370} (1992) 69--104}. [Addendum: Nucl.Phys.B 375, 501 (1992)].

\bibitem{Silvestrini:2018dos}
L.~Silvestrini and M.~Valli, {\em {Model-independent Bounds on the Standard
  Model Effective Theory from Flavour Physics},}
  \href{https://dx.doi.org/10.1016/j.physletb.2019.135062}{Phys.\  Lett.\  B
  {\bfseries 799} (2019) 135062} {\ttfamily
  [\href{https://arxiv.org/abs/1812.10913}{arXiv:1812.10913}]}.

\bibitem{Bagger:1997gg}
J.~A.~Bagger, K.~T.~Matchev, and R.-J.~Zhang, {\em {QCD corrections to flavor
  changing neutral currents in the supersymmetric standard model},}
  \href{https://dx.doi.org/10.1016/S0370-2693(97)00920-9}{Phys.\  Lett.\  B
  {\bfseries 412} (1997) 77--85} {\ttfamily
  [\href{https://arxiv.org/abs/hep-ph/9707225}{hep-ph/9707225}]}.

\bibitem{Lee:1998qq}
K.~Y.~Lee and J.~C.~Lee, {\em {$B$ - $\bar{B}$ mixings and rare $B$ decays in
  the top flavor model},}
  \href{https://dx.doi.org/10.1103/PhysRevD.58.115001}{Phys.\  Rev.\  D
  {\bfseries 58} (1998) 115001} {\ttfamily
  [\href{https://arxiv.org/abs/hep-ph/9803248}{hep-ph/9803248}]}.

\bibitem{Choudhury:2004bh}
S.~R.~Choudhury, N.~Gaur, A.~Goyal, and N.~Mahajan, {\em {$B_d$ - $\bar{B}_d$
  mass difference in little Higgs model},}
  \href{https://dx.doi.org/10.1016/j.physletb.2004.09.051}{Phys.\  Lett.\  B
  {\bfseries 601} (2004) 164--170} {\ttfamily
  [\href{https://arxiv.org/abs/hep-ph/0407050}{hep-ph/0407050}]}.

\bibitem{Grinstein:1990tj}
B.~Grinstein, R.~P.~Springer, and M.~B.~Wise, {\em {Strong Interaction Effects
  in Weak Radiative $\bar{B}$ Meson Decay},}
  \href{https://dx.doi.org/10.1016/0550-3213(90)90350-M}{Nucl.\  Phys.\  B
  {\bfseries 339} (1990) 269--309}.

\bibitem{Buras:1998raa}
A.~J.~Buras in {\em {Les Houches Summer School in Theoretical Physics, Session
  68: Probing the Standard Model of Particle Interactions}}, pp.~281--539.
\newblock 1998.
\newblock {\ttfamily
  \href{https://arxiv.org/abs/hep-ph/9806471}{hep-ph/9806471}}.

\bibitem{Ciuchini:1993fk}
M.~Ciuchini, E.~Franco, L.~Reina, and L.~Silvestrini, {\em {Leading order QCD
  corrections to $b \to s \gamma$ and $b \to s g$ decays in three
  regularization schemes},}
  \href{https://dx.doi.org/10.1016/0550-3213(94)90223-2}{Nucl.\  Phys.\  B
  {\bfseries 421} (1994) 41--64} {\ttfamily
  [\href{https://arxiv.org/abs/hep-ph/9311357}{hep-ph/9311357}]}.

\bibitem{Buras:1993xp}
A.~J.~Buras, M.~Misiak, M.~Munz, and S.~Pokorski, {\em {Theoretical
  uncertainties and phenomenological aspects of $B \to X_s \gamma$ decay},}
  \href{https://dx.doi.org/10.1016/0550-3213(94)90299-2}{Nucl.\  Phys.\  B
  {\bfseries 424} (1994) 374--398} {\ttfamily
  [\href{https://arxiv.org/abs/hep-ph/9311345}{hep-ph/9311345}]}.

\bibitem{Misiak:2015xwa}
M.~Misiak {\em et~al.}, {\em {Updated NNLO QCD predictions for the weak
  radiative B-meson decays},}
  \href{https://dx.doi.org/10.1103/PhysRevLett.114.221801}{Phys.\  Rev.\
  Lett.\  {\bfseries 114} (2015) 221801} {\ttfamily
  [\href{https://arxiv.org/abs/1503.01789}{arXiv:1503.01789}]}.

\bibitem{Endo:2017ums}
M.~Endo, {\em et al.}, {\em {Gluino-mediated electroweak penguin with
  flavor-violating trilinear couplings},}
  \href{https://dx.doi.org/10.1007/JHEP04(2018)019}{JHEP {\bfseries 04} (2018)
  019} {\ttfamily [\href{https://arxiv.org/abs/1712.04959}{arXiv:1712.04959}]}.

\end{thebibliography}\endgroup

\end{document}